\documentstyle[12pt]{article}
\begin{document}
\title{TOWARDS A UNIFIED DESCRIPTION OF THE FUNDAMENTAL INTERACTIONS}
\author{B.G. Sidharth\\ Centre for Applicable Mathematics \& Computer Sciences\\
B.M. Birla Science Centre, Hyderabad 500 063}
\date{}
\maketitle
\footnotetext{Email:birlasc@hd1.vsnl.net.in}
\begin{abstract}
We consider a recent successful model of leptons as Kerr-Newman type Black Holes
in a Quantum Mechanical context. The model leads to a cosmology which
predicts an ever expanding accelerating universe with decreasing density and
to the conclusion that at Compton wavelength scales, electrons would exhibit
low dimensionality, both of which conclusions have been verified by several
independent experiments and observations very recently. In this preliminary communication
we indicate how the above model describes the quarks' fractional charges,
handedness and masses (as any fundamental theory should) and could lead to a unified
description of the four fundamental interactions.
\end{abstract}
\section{Introduction}
In previous communications\cite{r1,r2,r3} a model for leptons as Kerr-Newman type
Black Holes was developed, and it was pointed out that several hitherto
inexplicable features turned out to have a natural explanation, for
example the quantum of charge, the electromagnetic-gravitational interaction
ratio, the handedness of the neutrino and so on. We briefly recapitulate some
relevant facts.\\
As is well known, the Kerr-Newman metric has the electron's anomalous $g = 2$
factor built in, apart from giving the correct field of the particle\cite{r4}.
In the model referred to, the horizon of the particle Black Hole was taken
to be at the Compton wavelength, within which, as is well known, the so called negative
energy components $\chi$ of the Dirac four spinor dominate. There is ofcourse,
in this case, a naked singularity, but this is shielded by Zitterbewegung effects
which are symptomatic of the fact, that in Quantum Mechanics we cannot have
arbitrarily small space-time intervals unlike in classical theory.  It was shown that
the fact that, under spatial reflections $\chi \to -\chi$ while $\phi \to
\phi$ where $\phi$ represents the positive energy components of the Dirac
spinor leads to the emergence of the electromagnetic field owing to the fact
that $\chi$ shows up as a tensor density of weight $N = 1$, so that its
derivative has the following behaviour:
\begin{equation}
\frac{\partial \chi}{\partial x^\mu} \to \frac{1}{\hbar} [\hbar \frac{\partial}
{\partial x^\mu} - NA^\mu] \chi\label{e1}
\end{equation}
where from (\ref{e1}), we have,
\begin{equation}
A^\mu = \hbar \Gamma_\sigma^{\mu \sigma} = \hbar \frac{\partial}{\partial x^\mu}
log (\sqrt{|g|})\label{e2}
\end{equation}
It was pointed out that this was how the double valued spin half of Quantum
Mechanics could enter a general relativistic formulation. Infact in a linearized
theory, we get (cf.ref.\cite{r1,r2,r3})
\begin{equation}
g_{\mu v} = \eta_{\mu v} + h_{\mu v}, h_{\mu v} = \int \frac{4T_{\mu v}
(t - |\vec x - \vec x'|,\vec x')}{|\vec x - \vec x'|} d^3 x'\label{e3}
\end{equation}
in the usual notation. As is well known these lead to the relations, in
geometrized units\cite{r5}
\begin{equation}
m = \int T^{oo} d^3 x\label{e4}
\end{equation}
\begin{equation}
S_k = \int \epsilon_{klm} x^l T^{mo} d^3 x\label{e5}
\end{equation}
It was shown that for the Compton wavelength boundary referred to earlier,
one recovers from equation (\ref{e5}) the spin $S_k = \frac{h}{2}$ of the
electron while from (\ref{e3}) and (\ref{e4}) we get the usual gravitational
potential,
\begin{equation}
\Phi = - \frac{1}{2} (g^{oo} - \eta^{oo}) = - \frac{m}{r} + 0 (\frac{1}{r^3})\label{e6}
\end{equation}
It was also shown that from equation (\ref{e2}) one could get
\begin{equation}
\frac{ee'}{r} = A_0 \approx \frac{2\hbar G}{r}\int \eta^{\mu \nu} \frac{d}{d\tau}T_{\mu \nu} d^3
x'\label{e7}
\end{equation}
where $e'$ is the test charge.\\
From (\ref{e7}), using (\ref{e6}), we can deduce that (cf.ref.\cite{r1} for
details),
\begin{equation}
\frac{ee'}{r} = A_0 \approx 2m\int \eta^{\imath j} \frac{T_{\imath j}}{r}
d^3x'\label{e8}
\end{equation}
whence, as the test charge $e' = e,$
\begin{equation}
e^2 \approx m^2\label{e9}
\end{equation}
or
\begin{equation}
m \sim 10^{-33}cm\label{e10}
\end{equation}
which is the Planck mass. In the usual physical units, (\ref{e9}) and
(\ref{e10}) become, respectively,
\begin{equation}
e^2 \approx Gm^2\label{e11}
\end{equation}
\begin{equation}
m \sim 10^{-5}g\label{e12}
\end{equation}
The content of (\ref{e11}) and (\ref{e12}) is that the electromagnetic
and gravitational interactions become equal at the Planck mass. For
elementary particles with mass $m_p \sim 10^{-20}m,$ (\ref{e11}) gives,
the well known relation,
\begin{equation}
\frac{e^2}{Gm^2_p} \sim 10^{40}\label{e13}
\end{equation}
Finally it was also shown that by the usual method, one could consistently get back
the Kerr-Newman metric from the above considerations, which was the original
inspiration(cf.ref.\cite{r6}).\\
Thus we have a reconciliation between electromagnetism and gravitation.
Neutrinos are also described by the above model. In this case as pointed
out, because they have vanishingly small mass, they have a very large
Compton wavelength so that at our usual spatial scales we encounter predominantly
the negative energy components with the peculiar handedness (cf.ref.\cite{r1}
for details).It was shown in ref.\cite{r1} that if on the other hand
instead of considering distances $>>$
the electron Compton wavelength we consider the distances of the order of the
Compton wavelength itself (\ref{e3}) leads to a QCD type potential,
\begin{eqnarray}
4 \int \frac{T_{\mu \nu} (t,\vec x')}{|\vec x - \vec x' |} d^3 x' +
(\mbox terms \quad independent \quad of \quad \vec x), \nonumber \\
+ 2 \int \frac{d^2}{dt^2} T_{\mu \nu} (t,\vec x')\cdot |\vec x - \vec x' |
d^3 x' + 0 (| \vec x - \vec x' |^2) \propto - \frac{\propto}{r} + \beta r\label{e14}
\end{eqnarray}
The above considerations immediately lead us to consider the possibility of describing
weak interactions on the one hand and quarks on the other
in the above model.\\
We will now indicate how it is possible to do so.
\section{Strong Interactions}
Let us start with the electrostatic potential given in equations (\ref{e7})
and (\ref{e8}).\\
We will first show how the characteristic and puzzling $\frac{1}{3}$ and
$\frac{2}{3}$ charges of the quarks emerge.\\
For this we first note that the electron's spin half which is correctly
described in the above model of the Kerr-Newman Black Hole, outside the
Compton wavelength  automatically implies three spatial dimensions\cite{r5,r7}.
This is no longer true as we
approach the Compton wavelength in which case we deal with low space dimensionality. This indeed has been already
observed in experiments with nanotubes\cite{r8,r9}. In other words for the Kerr-Newman
Fermions spatially confined to distances of the order of their Compton
wavelength or less, we actually have to consider two and one spatial
dimensionality.\\
Using now the well known fact\cite{r10} that each of the $T_{\imath j}$ in
(\ref{e7}) or (\ref{e8}), is given by $\frac{1}{3} \epsilon, \epsilon$
being the energy density, it follows from (\ref{e8}) that the
particle would have the charge $\frac{2}{3} e$ or $\frac{1}{3}e$, as in the
case of quarks. Moreover, as noted earlier (cf.ref.1 also), because we are
at the Compton wavelength scale, we encounter predominantly the components
$\chi$ of the Dirac wave function, with opposite parity. So, as with neutrinos,
this would mean that the quarks would display helicity, which indeed is true:
As is well known, in the $V-A$ theory, the neutrinos and relativistic quarks
are lefthanded while the corresponding anti-particles are right handed (brought
out by the small Cabibo angle). This also automatically implies that these fractionally
charged particles cannot be observed individually because they are by their
very nature spatially confined. This is also expressed by the confining part
of the QCD potential (\ref{e14}). We come to this aspect now.\\
Let us consider the QCD type potential (\ref{e14}). To facilitate comparison
with the standard literature\cite{r11}, we multiply the left hand expression
by $\frac{1}{m}$ (owing to the usual factor $\frac{\hbar^2}{2m})$ and also
go over to natural units $c = \hbar = 1$ momentarily. The potential then
becomes,

\begin{equation}
\frac{4}{m} \int \frac{T_{\mu v}}{r} d^3 x + 2m \int T_{\mu v} r d^3x \equiv
-\frac{\propto}{r} + \beta r\label{e15}
\end{equation}
Owing to (\ref{e4}), $\propto \sim O(\ref{e1})$ and $\beta \sim O(m^2),$
where $m$ is the mass of the quark. This is indeed the case for the QCD
potential (cf.ref.\cite{r11}). Interestingly, as a check, one can verify
that, as the Compton wavelength distance $r \sim \frac{1}{m}$ (in natural
units), the energy given by (\ref{e15}) $\sim O(m)$, as it should be.\\
Thus both the fractional quark charges (and handedness) and their masses are seen to arise from
this formulation.\\
To proceed further we consider (\ref{e8}) (still remaining in natural units):
\begin{equation}
\frac{e^2}{r} = 2Gm_e \int \eta^{\mu v}\frac{T_{\mu v}}{r}d^3x\label{e16}
\end{equation}
where at scales greater than the electron Compton wavelength, $m_e$ is the
electron mass. At the scale of quarks we have the fractional charge and
$e^2$ goes over to $\frac{e^2}{10} \approx \frac{1}{1370} \sim
10^{-3}$.\\
So we get from (\ref{e16})
$$\frac{10^{-3}}{r} = 2Gm_e \int \eta^{\mu v} \frac{T_{\mu v}}{r} d^3 x$$
or,
$$\frac{\propto}{r} \sim \frac{1}{r} \approx 2G.10^3 m_e \int \eta^{\mu v}
\frac{T_{\mu v}}{r} d^3x$$
Comparison with the QCD potential and (\ref{e16}) shows that the now fractionally charged
Kerr-Newman fermion, viz the quark has a mass $\sim 10^3 m_e$, which is
correct.\\
If the scale is such that we do not go into fractional charges, we
get from (\ref{e16}), instead, the mass of the intermediary particle as
$274 m_e,$ which is the pion mass.\\
All this is ofcourse completely consistent with the physics of strong
interactions.
\section{Weak Interactions}
It has already been noted that in the formulation of leptons as Kerr-Newman
Black Holes, for Neutrinos, which have vanishingly small mass, if at all, so
that their Compton wavelength is infinite or very large, we encounter
predominantly the negative energy components of the Dirac spinor. It was
shown in, for example, ref.\cite{r1}, that this explains their characteristic
helicity and two component character. One should expect that from the
above considerations, we should be able to explain weak interactions also.\\
Even in the early days leading to the electro-weak theory,\cite{r12,r13}, it
was realized, that with the weak coupling constant set equal to the electromagnetic
coupling constant and with a massive intermediary particle $m_w \sim 100 m_p,$
where $m_p$ is the proton mass, we get the Fermi local weak coupling constant,\\
\begin{equation}
G_w = g^2/m^2_w \approx \frac{10^{-5}}{m^2_p}gm^{-2}\label{e17}
\end{equation}
This is also the content of our argument: We propose to show now that (\ref{e17})
is consistent with (\ref{e11}), just as at the
mass scale of electrons or protons, we get (\ref{e13}) from (\ref{e11}).
However, it should be borne in mind that now (\ref{e11}) is not an adhoc
experimental result, but rather follows from our model, from equations
(\ref{e1}) to (\ref{e3}).\\
Our starting point is equation (\ref{e11}), which we rewrite, as
$$\frac{e^2\times 10^{19} \times 10^8}{10^4 m^2_p} \approx \frac{10^{40}\times
10^{19}}{10^4}\approx 10^{55},$$
remembering that $G \sim 10^{-8}$ and $e^2 \sim 10^{-19}$. From here we get,
$$\frac{g^2}{m_w^2} \approx 10^{43} gm^{-2}$$
with $g^2 \approx 10^{-1}$ and $m_w \approx 100 m_p$ which is
consistent with (\ref{e17}).
\section{Discussion and Conclusion}
The model of leptons as Kerr-Newman Black Holes, as discussed in Section 1
leads to a cosmology\cite{r2,r14,r15} in which the universe continues to
expand and accelerate with ever decreasing density. Interestingly this
has been confirmed by several independent recent observations\cite{r16,r17,r18}.
On the other hand the model also predicts that near Compton wavelength scales
electrons would display a neutrino type bosonization or low dimensionality\cite{r19}.
As pointed out earlier (cf.refs.\cite{r8,r9}), this has been confirmed in
the case of recently developed carbon nanotubes.\\
What we have shown above is that the Kerr-Newman Black Hole description
of the electron leads to a unification of gravitation and electromagnetism
as expressed by (\ref{e11}) or (\ref{e13}). It also gives a clue to the
peculiar fractional charges and also masses of the quarks and the QCD interaction,
as expressed by (\ref{e15}). Finally the model explains the handedness
of the neutrino and gives a clue to the origin of weak interactions, as
expressed by (\ref{e17}).\\
So at the heart of the matter is the description of the electron as a
Kerr-Newman Black Hole, valid at length scales greater than the Compton
wavelength. The other phenomena appear from here at different length (or
energy) scales.

\end{document}